\documentclass{appolb}
\usepackage{graphicx}

\begin{document}
\title{The Langevin approach for fission of heavy and super-heavy nuclei%
\thanks{Presented at the Zakopane Conference on Nuclear Physics, Zakopane, Poland, 28 August - 4 September 2022}
}
\author{F.A.Ivanyuk, S.V.Radionov
\address{Institute for Nuclear Research, Kyiv, Ukraine}
\\[3mm]
{C.Ishizuka, S.Chiba 
\address{Tokyo Institute of Technology, Tokyo, Japan}
}}
\maketitle
\begin{abstract}
In this contribution, we present the main relations of the Langevin approach to the description of fission or fusion-fission reactions. The results of Langevin calculations are shown for the mass distributions of fission fragments of super-heavy elements and used for the investigation of memory effects in nuclear fission.

\end{abstract}

\section{Introduction}
We describe the nuclear fission process by the four-dimensional set of the Langevin equations for the shape degrees of freedom with the shape given by the two-center shell model (TCSM) shape parametrization.
The potential energy is calculated within the macroscopic-microscopic method.
The collective mass, $M$,  and friction, $\gamma$, tensors are defined in macroscopic (Werner-Wheller and wall-and-window formula) or microscopic (linear response theory) approaches.

We start calculations from the ground state shape with zero collective velocities and solve equations until the neck radius of the nucleus turns zero (scission point). At the scission point, the solutions of Langevin equations supply complete information about the system, its shape, excitation energy, and collective velocities.
This information makes it possible to calculate the mass distributions, the total kinetic energy, and the excitation energies of fission fragments.
The results of numerous previous calculations are in reasonable agreement with the available experimental data.

Below in this contribution, we present the calculated results for the mass distributions of super-heavy nuclei and clarify the impact of memory effects on the fission width of heavy nuclei.

The physics of super-heavy elements (SHE) has a long history. The existence
of the ``island of stability'' was predicted at the end of the 1960s \cite{1}. Nevertheless, it took almost 30
years until the alpha-decay of the element with Z=114 was observed
experimentally at Flerov Nuclear Reactions Laboratory in Dubna \cite{5}.

With the development of experimental facility, it became possible not only to fix the fact of formation of SHE, but examine their properties. One of the first property of interest -- the process of fission of SHEs. For the successful planning and carrying out of experiments, it is crucial to understand what kind of fission fragments mass distribution (FFMD) one should expect in the result of the fission of SHEs. The two double magic nuclei $^{132}$Sn and $^{208}$Pb may contribute.
Both have the shell correction in the ground state of the same magnitude.

In order to clarify what kind of FFMD one could expect in the fission of SHEs,
we have carried out the calculations of FFMD for a number of SHEs. The results are given in Section 3.

Another problem we address in this contribution is the influence of memory effects on the probability of the fission process.
Commonly one uses the Markovian approximation to Langevin approach in which all quantities are defined at the same moment. This approximation provides reasonable results, but its accuracy is not well established. In publications, one can find statements that the memory effects have a significant influence on the fusion or fission
processes and the statements that memory effects are very small.

To clarify this uncertainty, we have calculated the fission width using the Langevin approach with memory effects included in a wide range of important parameters: the excitation energy $E^*$ of the system, the damping parameter $\eta$, the relaxation time $\tau$.
The details and results of the calculations are given in Section 4.

\section{The Langevin approach for the fission process}
Within the Langevin approach, the fission process is
described by solving the equations for the time evolution of
the shape of nuclear surface of the fissioning system. For the shape
parametrization, we use that of the two-center shell model (TCSM)
\cite{16}
with 4 deformation parameters $q_{\mu} =z_0/R_0, \delta_1, \delta_2, \alpha$.
Here z$_0/R_0$ refers to the distance
between the centers of left and right oscillator potentials, $R_{0}$ being the radius of spherical nucleus with the mass
number A. The parameters $\delta _i$ describe the
deformation of the right and left fragment tips. The fourth parameter
$\alpha $ is the mass asymmetry and the fifth parameter of the TCSM shape
parametrization $\epsilon $ was kept constant, $\epsilon $=0.35, in
all our calculations.

The first-order differential equations (Langevin equations) for the time
dependence of collective variables $q_{\mu }$ and the conjugated momenta
$p_{\mu }$ are:
\begin{eqnarray}
\label{lange}
\frac{dq_\mu}{dt}&=&\left(m^{-1} \right)_{\mu \nu} p_\nu , \\
\frac{dp_\mu}{dt}&=&-\frac{\partial F(q,T)}{\partial q_\mu} - \frac{1}{2}\frac{\partial m^{-1} _{\nu \sigma} }{\partial q_\mu} p_\nu p_\sigma
 -\gamma_{\mu \nu} m^{-1}_{\nu \sigma} p_\sigma + R_\mu (t). \nonumber
\end{eqnarray}
In Eqs. (\ref{lange}) the $F(q, T)$ is the
temperature-dependent free energy of the system, and $\gamma _{\mu \nu }$
and (m$^{-1})_{\mu \nu }$ are the friction and inverse of mass tensors.

The free energy $F(q, T)$ is calculated within the shell correction method.
The single particle energies are calculated with the deformed Woods-Saxon potential fitted to the mentioned above TCSM  shapes.

The collective inertia tensor $m_{\mu \nu }$ is calculated by the Werner-Wheeler approximation
and for the friction tensor $\gamma _{\mu \nu }$ we used the wall-and-window formula.
The random force $R_{\mu}(t)$ is the product of the temperature-depen\-dent strength factors
g$_{\mu \nu }$ and the white noise $\xi_{\nu}$(t), $R_{\mu}(t)=g_{\mu \nu }\xi_{\nu }$(t). 
The factors g$_{\mu \nu }$ are related to the temperature
and friction tensor via the Einstein relation,
\begin{equation}\label{Einstein}
g_{\mu \sigma } g_{\sigma \nu } =T\gamma _{\mu \nu}
\end{equation}
The temperature T is kept constant, $aT^2=E^*$, or adjusted to the local excitation energy on each step of integration by the relation,
\begin{equation}\label{temper}
aT^2=E^*-p^2(t)/2M-[E_{pot}(q)-E_{pot}(q_{gs})].
\end{equation}
Here $q_{gs}$ is the ground state deformation.
More details are given in our earlier publications \cite{24,25,26,14}.

Initially, the momenta $p_{\mu }$ are set to zero, and calculations are
started from the ground state deformation. Such calculations are continued
until the trajectories reach the "scission point", defined as the
point in deformation space where the neck radius turns zero. 
\section{Fission fragments mass distributions of super-heavy nuclei}
\begin{figure}[htb]
\centerline{%
\includegraphics[width=12.5cm]{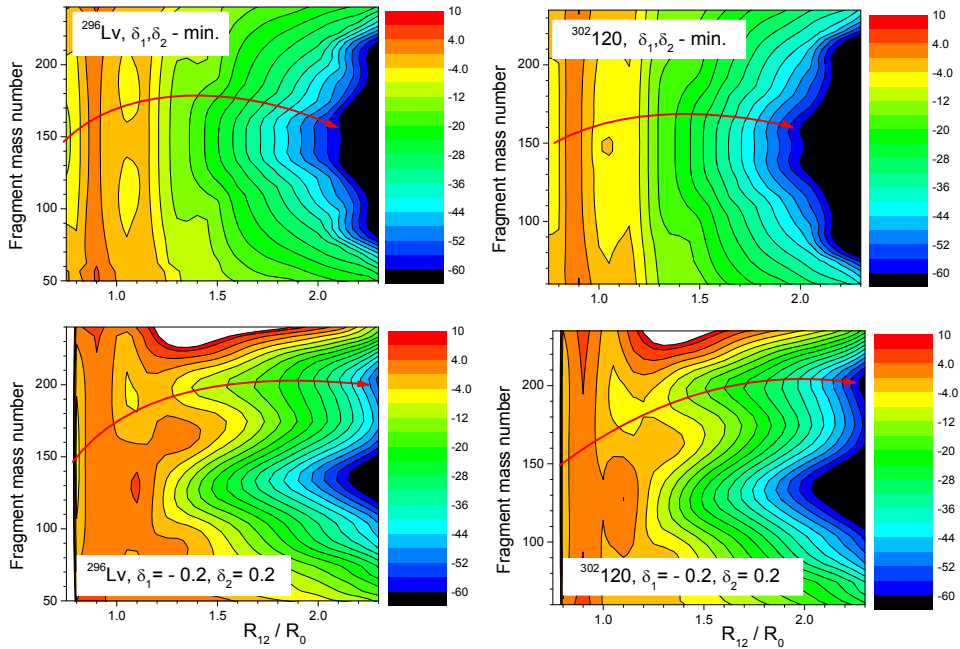}}
\caption{(top) The potential energy of $^{296}Lv$ and $^{302}120$ at $T=0$ minimized with respect to deformation parameters $\delta_1$ and $\delta_2$ (bottom), and at fixed values $\delta_1=-0.2$ and $\delta_2=0.2$.}
\label{fig1}
\end{figure}

In order to understand what kind of mass distributions one can expect from the solution of Langevin equations for super-heavy nuclei, we looked first at the potential energy of fissioning nuclei.
Fig. \ref{fig1} shows the potential energy E$_{def}$ of nuclei $^{296}$Lv and $^{302}$120
at zero temperature as a function of elongation (the distance R$_{12 }$ between
the centers of mass of left and right parts of a nucleus) and the mass asymmetry (fragment mass number). In the top part of
Fig. \ref{fig1} the energy was minimized with respect to the deformation parameters
$\delta _{1}$ and $\delta _{2}$. One sees the bottom of
potential energy leading to almost symmetric mass splitting. There is also a
hint on the mass asymmetric valley at $A_F$ close to $A_F$=208.

If the trajectories followed the bottom of potential energy, the mass distributions
would be symmetric. However, it is well known that the trajectories may deviate substantially from the bottom
of the potential valley due to dynamic effects. We calculate the trajectories in four-dimensional
deformation space. In this space, the local minima could lead
away from the bottom of the potential valley. An example is shown in the bottom
part of Fig. \ref{fig1}. Here we show the potential energy for fixed $\delta _1$=
- 0.2 and $\delta _2$=0.2. One clearly sees another valley, leading to strongly mass asymmetric splitting.

\begin{figure}[htb]
\centerline{%
\includegraphics[width=12.5cm]{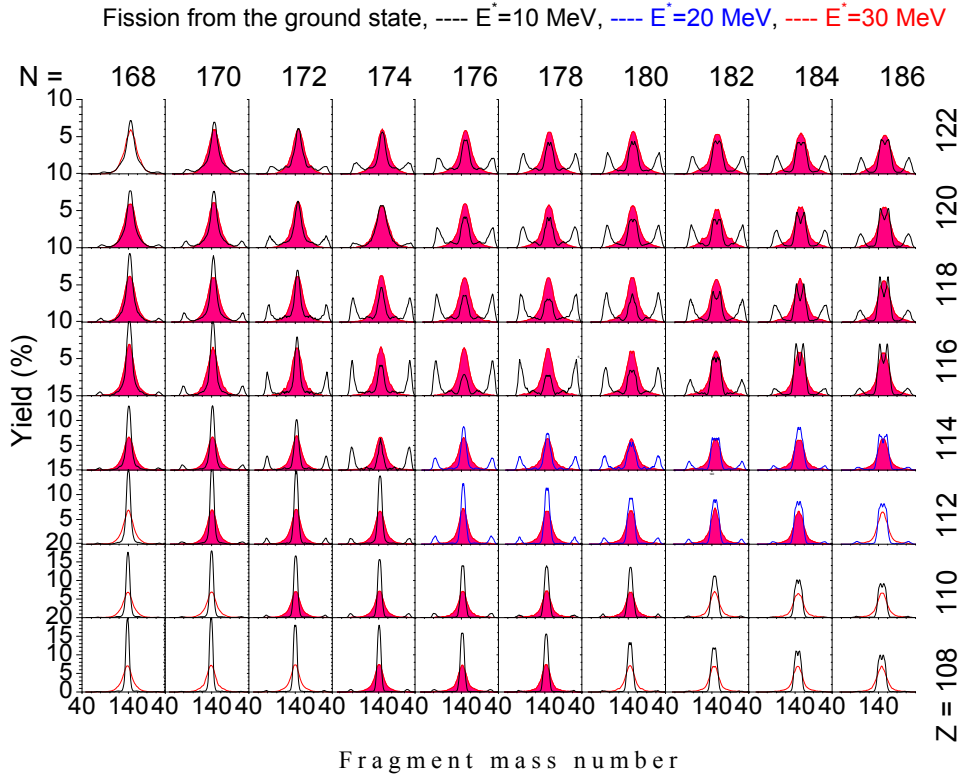}}
\caption{The fission fragment mass distributions  of super-heavy nuclei from $^{276}$Hs to $^{308}$122 calculated for the excitation energies $E^*$=10, 20 and 30 MeV as a function of the fragment mass number}
\label{fig2}
\end{figure}

In Fig. \ref{fig2}, we show the fission fragment mass distributions of super-heavy
nuclei from $^{276}$Hs to $^{308}$122 as a function of fragment mass number
$A_F$. The FFMDs of nuclei from $^{276}$Cn to $^{308}$122 have three or
four peak structures.
The main component is the symmetric peak, split
into two components in some isotopes. The peaks of lighter fragments are located around
$A_F$=140.

One can also see the strongly asymmetric peak at the mass number close to
$A_F$=208. The strength of the (almost) symmetric and asymmetric
components in FFMD of SHEs depends on the proton and neutron numbers of the
compound nucleus. For $^{276}$Cn, the contribution of a strongly asymmetric
peak is tiny. This contribution becomes larger for more heavy SHE. In
some elements of SHEs with $Z=$116-122, the symmetric and mass-asymmetric peaks are of the
same magnitude. More details can be found in \cite{SHEs}.

The similar strongly mass-asymmetric peaks in FFMD of SHEs were also found recently in \cite{pasha} within the Langevin approach with the so call Fourier shape parametrization.

\section{The memory effects in nuclear fission}
\begin{figure}[htb]
\centerline{%
\includegraphics[width=6.0cm,height=4.2cm]{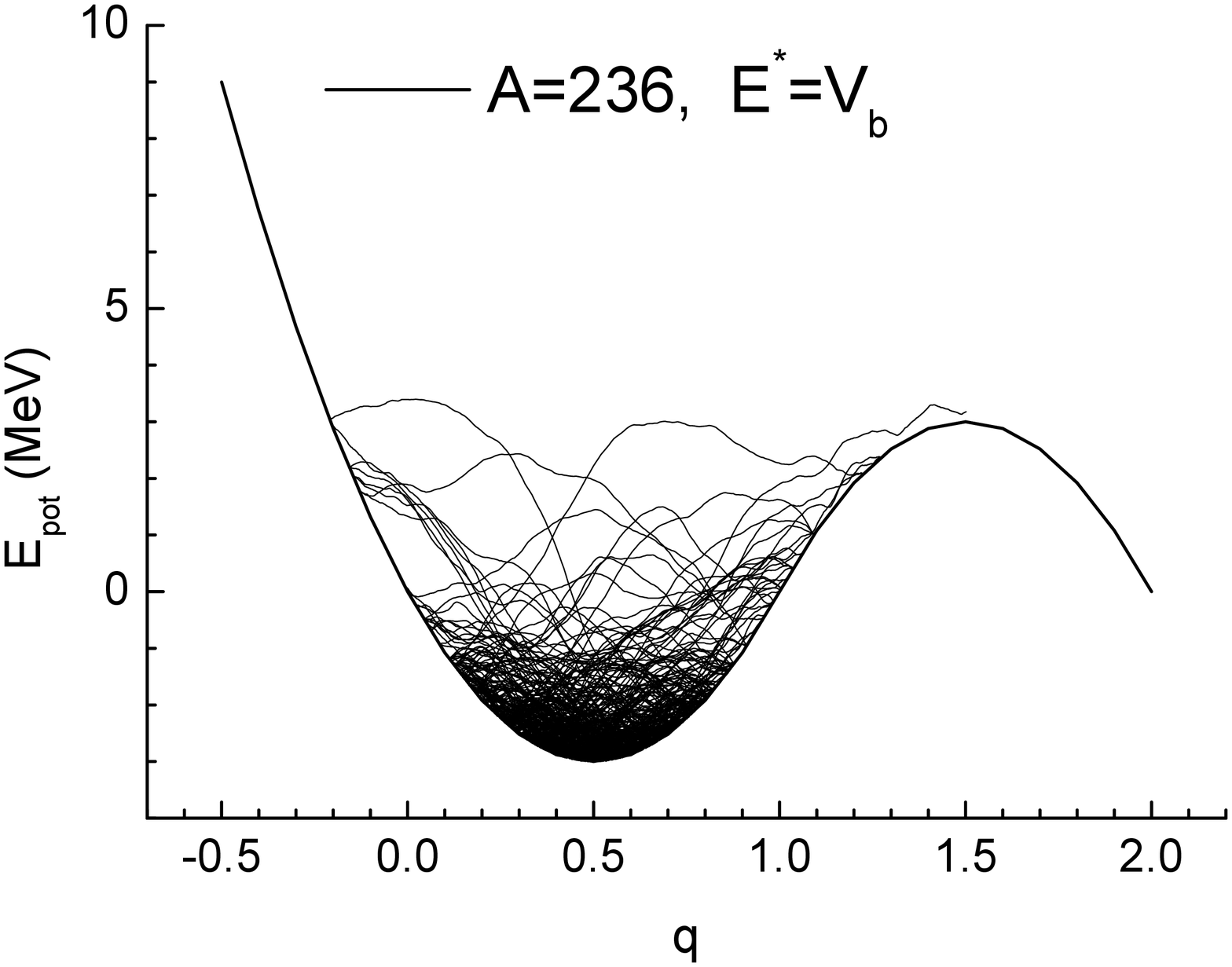}
\includegraphics[width=6.0cm]{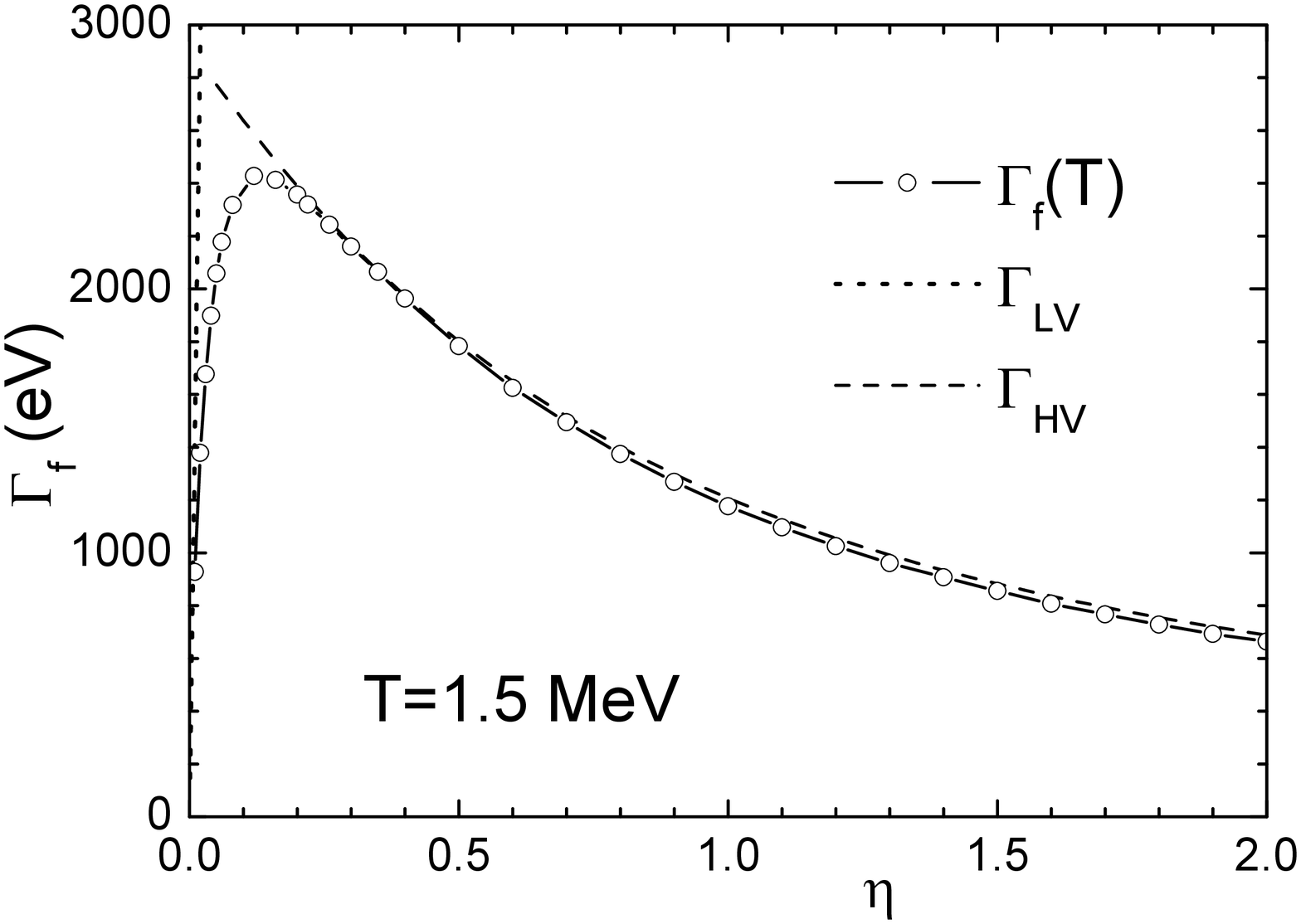}}
\caption{(left) The two-parabolic potential (\ref{poten}) and few examples of the dynamical trajectories.
(right) The fission width as the solution of Eqs.~(\ref{lange}, \ref{poten}, \ref{decay}) calculated at constant temperature (open dots), and the Kramers approximations (\ref{GammaK}) for high and low damping limits. }
\label{fig3}
\end{figure}

In order to investigate the role of memory effects in nuclear fission, we exploit a simple one-dimensional model with the potential energy given by the two-parabolic potential (Kramers potential), see Fig. \ref{fig3}.
\begin{equation}\label{poten}
{\rm E}_{pot}(q)=2V_bq(q-q_0)/q_0^2,\, 0<q<q_0; 2V_b(q-q_0)(2q_0-q)q_0^2,\,q_0<q<2q_0.
\end{equation}
The potential (\ref{poten}) depends on two parameters, the barrier height $V_b$ and the barrier width $q_0$. We have fixed the barrier height $V_b$ = 6 MeV, which is close to the value of the fission barrier of actinide nuclei. The width of the barrier is somewhat uncertain. It depends on the definition of the collective coordinate $q$ and the model for the potential energy. For simplicity, we have put here $q_0=1.0$.

For the potential (\ref{poten}) one can define the stiffness $C=d^2E_{pot}/dq^2$ and the frequency of harmonic vibrations $\omega_0=\sqrt{C/M}$. In the present work, we fix $\hbar\omega_0$ =1.0 MeV, which is close to the frequency of collective vibrations calculated for $^{224}$Th in \cite{hiry} within the microscopic linear response theory. Then, for the mass parameter we will have the deformation and temperature-independent value,
\begin{equation}\label{mass}
M=C / \omega_0^2 = 4 V_b / (\omega_0^2q_0^2).
\end{equation}
For the friction coefficient $\bar\gamma$ we use a slightly modified approximation of \cite{hiry}, 
\begin{equation}\label{beta}
\bar\gamma/M=0.6(T^2+\hbar^2\omega_0^2/\pi^2))/(1+T^2/40).
\end{equation}
For the temperature, we consider here two options: constant temperature regime and constant energy regime.
In a constant temperature regime, the temperature is time-independent, related to the initial excitation energy $E^*$ by the Fermi-gas relation, $a T^2=E^*$,
where $a$ is the level density parameter of T\"oke and Swiatecki \cite{toke}.
The fission width calculated in a constant temperature regime will be denoted as $\Gamma_f(T)$.

At small excitations, the temperature varies with deformation and time, and there is no reason to consider it constant. So, it should be adjusted to the local excitation energy on each integration step by the relation (\ref{temper}).
Correspondingly, fission width calculated in a constant energy regime is denoted as $\Gamma_f(E)$.

The fission width, $\Gamma_f$, is defined assuming the exponential decay of the number of  "particles" in the potential well,
\begin{equation}\label{decay}
P(t)=e^{-\Gamma_f t/\hbar} \to \Gamma_f=-\hbar \ln[P(t)]/t.
\end{equation}
By solving the Langevin equations one will get the set of time moments
$t_b$, at which some trajectories would cross the barrier. From this information, one can find the probability $P(t)$ and the fission width $\Gamma_f$, see \cite{memory}.

The Markovian fission width $\Gamma_f(T)$ calculated by Eqs. (\ref{lange}, \ref{poten}, \ref{decay}) is plotted  as function of the damping parameter $\eta$  in the right part of Fig.~\ref{fig3}.
To present the results in a broader range of parameters, the damping parameter $\eta\equiv\bar\gamma/2M\omega_0$ in these calculations was considered as a free parameter.

For the comparison, in Fig.3 we also show the Kramers decay width $\Gamma_{HV}, \Gamma_{LV} $ in limits of high and low viscosity (friction) \cite{kramers},
\begin{equation}\label{GammaK}
\Gamma_{HV}=\frac{\hbar\omega_0}{2\pi}e^{-V_b/T}(\sqrt{1+\eta^2}-\eta)\,,\quad
\Gamma_{LV}=\frac{\hbar\bar\gamma}{M}\frac{V_b}{T}e^{-V_b/T}.
\end{equation}

As one can see, the dependence of $\Gamma_f(T)$ on $\eta$ is rather complicated. The fission width $\Gamma_f(T)$ {\it grows} as function of $\eta$ in low damping region ($\eta < 0.1$). For $\eta > 0.2$,  the fission width $\Gamma_f(T)$ {\it decreases} as function of $\eta$.

In nuclear systems, the Markovian assumption is often too restrictive. We thus have to generalize the above Langevin equations to allow for finite memory effects. They read as
 \cite{abe-san},
\begin{eqnarray}\label{lange2}
dq/dt&=&p(t)/M ,\\
\frac{dp}{dt}&=&-\frac{\partial E_{pot}}{\partial q} - \int_0^t dt^{\prime}\gamma(t-t^{\prime}) p(t^{\prime}) /M + \zeta\,,\,
\quad\gamma(t-t^{\prime})\equiv\bar\gamma e^{-\frac{t-t^{\prime}}{\tau}}\slash\tau\,,\nonumber
\end{eqnarray}
 where $\tau$ is the memory (or relaxation) time. The extension consists in allowing the friction to have a memory time, i.e., the friction reacts on past stages of the system, what is called a retarded friction.

The random numbers $\zeta$ in (\ref{lange2}) are the normally distributed random numbers with the properties $<\zeta(t)>=0$, $<\zeta(t)\zeta(t^{\prime})>=T\gamma (t-t^{\prime})$. In the limit $\omega_0\tau <<1$, one recovers the Markovian limit of nuclear fission dynamics, i.e., when the friction force is simply given by $\gamma \dot q(t)$.
The random numbers $\zeta(t)$ in (\ref{lange2}) satisfy the equation
\begin{equation}\label{zetat}
d\zeta(t)/dt=- \zeta(t) / \tau +R(t)/\tau\,,
\end{equation}
and are used in the description of the so-called Ornstein-Uhlenbeck processes.

In the top part of Fig.~4 the calculated fission width $\Gamma_f(E)$ is shown as a function of the damping parameter $\eta$   both for small and large excitation energies, $E^*$=10, 25 and 60 MeV, for few values of the relaxation time.
Besides $\tau=0$, we  choose in calculations below the two values of $\tau$, $\tau=5\cdot 10^{-22}$ sec and $\tau=10^{-21}$ sec.
\begin{figure}[htb]
\centerline{%
\includegraphics[width=12.5cm]{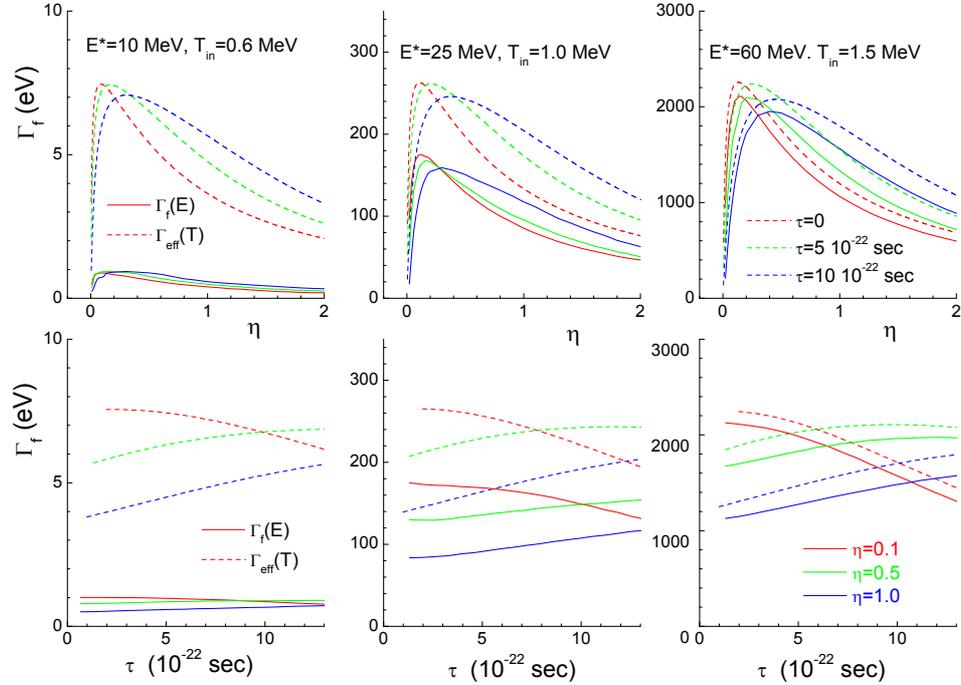}}
\caption{(top) The dependence of the fission width $\Gamma_f(E)$ (solid) and the approximation (\ref{gammaeff}) (dashed) on the damping parameter $\eta$ for few values of the relaxation time $\tau$, $\tau$=0, $\tau=5\cdot 10^{-22}$ sec,  $\tau=10^{-21}$ sec and the initial excitation energies $E_{in}^*$=10, 25 and  60 MeV.
(bottom) The dependence of the fission width $\Gamma_f(E)$ (solid) and the approximation (\ref{gammaeff}) (dashed) on the relaxation time $\tau$  for a few values of the damping parameter $\eta$, $\eta$=0.1, 0.5 and 1.0.}
\label{fig4}
\end{figure}

The results of Langevin calculations satisfying the energy conservation condition are shown in Fig.~4 by solid lines. The fission width $\Gamma_f(E)$ {\it grows} as a function of $\eta$ and {\it decreases} as a function of $\tau$ in low damping region. The tendency is the opposite in the high damping region; the fission width $\Gamma_f$ {\it falls} as a function of $\eta$ and {\it increases} as a function of $\tau$. Such dependence is common both for small and large excitation energies.

In the bottom part of Fig.~4, the fission width $\Gamma_f(E)$ (solid lines) is shown as a function of the relaxation time $\tau$ for a few fixed values of the damping parameter $\eta$. The bottom part of Fig.~4 confirms the above conclusion: the dependence of fission width $\Gamma_f$ on $\eta$ and $\tau$ is opposite in low and high damping regions.

For the comparison, we show by dashed lines in Fig.~4 the available analytical approximation for $\Gamma_f(T, \tau)$  \cite{abe-san,grote,lallouet},
\begin{equation}\label{gammaeff}
\frac{1}{\Gamma_{eff}}=\frac{1}{\Gamma_{LV}}+\frac{1}{\Gamma_{HV}}\,,\quad
\Gamma_{LV}(\tau)=\frac{\Gamma_{LV}(0)}{1+\omega_0^2\tau^2},\quad \Gamma_{HV}(\tau)=\frac{\hbar\lambda}{2\pi}e^{-V_b/T}\,,
\end{equation}
where $\lambda$ is the largest positive solution of the secular equation,
\begin{equation}\label{laluet}
\lambda^3+\lambda^2/\tau+(\bar\gamma/M\tau-\omega_0^2)\lambda-\omega_0^2/\tau=0\,.
\end{equation}
As can be seen, the results of Langevin calculations for $\Gamma_f(E)$ are smaller than the analytical estimate (\ref{gammaeff}) both in low and high damping limits. The ratio $\Gamma_f(E)/\Gamma_{eff}$ is close to 1 at $E^*$=60 MeV and close to 0.1 at $E^*$=10 MeV.

\section{Summary}
The calculated  mass distributions of fission fragments of super-heavy nuclei from $^{268}$Hs to $^{308}$122 demonstrate
a three-four peaks structure of mass distributions.
In light super-heavies, we see the dominant mass symmetric peak at $A_F\approx$ 140.
With increasing mass and charge numbers  of fissioning nuclei, the highly asymmetric peaks at
$A_H\approx$ 208  appears.
In $^{290-296}$Lv and $^{290-296}$Og, the three peaks in FFMD are approximately of the same magnitude at E*=10 MeV.

The investigation of memory effects in nuclear fission is carried out. The calculations presented here offer complete information on the dependence of fission probability on all essential parameters, the relaxation time $\tau$, the damping parameter $\eta$, and the excitation energy E*.

It turned out that the fission width $\Gamma_f(E)$ calculated under the constant energy requirement is generally smaller than that calculated in the constant temperature regime, $\Gamma_f(T)$, or the Bohr-Wheeler approximation.

The dependence of the fission width $\Gamma_f(E)$ on the relaxation time $\tau$ is very sensitive to the damping parameter $\eta$.  In the low viscosity region, the fission width $\Gamma_f(E)$ grows as a function of $\eta$ and decreases as a function of $\tau$. In the high-viscosity region, the tendency is the opposite. Such dependence is common both for small and large excitation energies.

{\bf Acknowledgements.} The authors are grateful to Prof. K.Pomorski for the valuable discussions and presentation of our results at the Zakopane Conference

\end{document}